\documentstyle[aps,prl,epsf]{revtex}

\newcommand{\nc}{\newcommand}
\nc{\beq}{\begin{equation}}
\nc{\eeq}{\end{equation}}
\nc{\beqa}{\begin{eqnarray}}
\nc{\eeqa}{\end{eqnarray}}
\nc{\lra}{\leftrightarrow}
\def\sfrac#1#2{{\textstyle\frac#1#2}}
\nc{\sss}{\scriptscriptstyle}
\nc{\PL}{P_{\sss L}}
\nc{\PR}{P_{\sss R}}
\nc{\lsim}{\mbox{\raisebox{-.6ex}{~$\stackrel{<}{\sim}$~}}}
\nc{\gsim}{\mbox{\raisebox{-.6ex}{~$\stackrel{>}{\sim}$~}}}

\def\dsl{\,\partial\!\!\!/}
\def\muL{{\mu_{\sss L}}}
\def\avg#1{{\left\langle #1 \right\rangle}}
\def\Gsph{\Gamma_{\rm sph}}

\begin{document}
\twocolumn[\hsize\textwidth\columnwidth\hsize\csname@twocolumnfalse%
\endcsname


\title{A New Source for Electroweak Baryogenesis in the MSSM}

\author{James M.~Cline}

\address{Physics Department, McGill University,
3600 University Street, Montr\'eal, Qu\'ebec, Canada H3A 2T8}
\author{Kimmo Kainulainen}
\address{\it  NORDITA, Blegdamsvej 17, DK-2100, Copenhagen \O , Denmark}

\maketitle

\begin{abstract} One of the most experimentally testable explanations
for the origin of the baryon asymmetry of the universe is that it
was created during the electroweak phase transition, in the minimal
supersymmetric standard model.  Previous efforts have focused on
the current for the difference of the two Higgsino fields, $H_1-H_2$,
as the source of biasing sphalerons to create the baryon asymmetry.  
We point out that the current for the orthogonal linear combination,
$H_1+H_2$, is larger by several orders of magnitude.  Although this
increases the efficiency of electroweak baryogenesis, we nevertheless
find that large CP-violating angles $\ge 0.15$ are required to
get a large enough baryon asymmetry. \\
\leftline{PACS: 98.80.Cq}\\
\end{abstract}

]

\noindent {\bf 1.}
A highly constrained proposal for the origin of baryonic matter in the
universe is electroweak baryogenesis in the minimal supersymmetric
standard model (MSSM) \cite{CN,HN,AOS,CQRVW,CJK,Rius}.  Unlike many other
baryogenesis mechanisms, this one has strong prospects for being falsified
in upcoming experiments, due to its need for some light exotic particles,
notably a right-handed top squark which is lighter than the top quark
\cite{phase_transition,CM}. 

The basic mechanism \cite{CKN} is intuitively clear: particles interact in
a CP-violating manner with bubble walls, which form during the first order
electroweak phase transition, when the temperature of the universe was
near $T = 100$ GeV.  This causes a buildup of a left-handed quark density 
in excess of that of the corresponding antiquarks, and an equal and
opposite right-handed asymmetry, so that there is initially no net baryon
number.  The left-handed quark asymmetry biases anomalous sphaleron
interactions, present within the standard model, to violate baryon number
preferentially to create a net quark density.  The resulting
baryon asymmetry of the universe (BAU) soon falls inside
the interiors of the expanding bubbles, where the sphaleron interactions
are shut off, and thus baryon number is safe from subsequent
sphaleron-induced relaxation to zero. 

Despite the simplicity of this idea, a quantitatively accurate
calculation is difficult to carry out.  Various approximations have
been made, leading to a variety of formalisms which give somewhat
conflicting predictions for the dependence of the BAU on the parameters
of the MSSM.  Although most authors agree on the diffusion equations
which govern the generation of the left-handed quark asymmetry, there
is less consensus about how to derive the source terms which appear in
these equations.

In a previous paper \cite{CJK}, we advocated an approach based on the
classical force on particles \cite{JPT,JKP} due to their spatially
varying masses as inside the bubble wall.  It is straightforward to
solve for this force, put it into the Boltzmann equations, and derive
corresponding diffusion equations.  No {\it ad hoc} assumptions are
needed; one only requires that the width of the wall be significantly
larger than thermal de Broglie wavelengths of particles in the plasma,
to justify an expansion in derivatives of the background Higgs field,
which constitutes the bubble wall.  Detailed calculations of the wall
width confirm that this is indeed a good approximation \cite{CMnote}.

One puzzling conflict between our earlier work \cite{CJK} and that of
others was that we derived a source which remains large when the ratio of
the two Higgs fields, $H_2(x)/H_1(x)$, is constant inside the bubble wall. 
Other authors \cite{HN,AOS,CQRVW} found sources proportional to the derivative
of this quantity. Careful analyses of the shape of the wall have shown
that in fact $H_2/H_1$ remains constant to within a part in $10^2-10^3$
\cite{MQS,CM}, so that the dependence on $d(H_2/H_1)/dx$,
if correct, would result in a large suppression of the generated BAU. 

In this Letter we explain the origin of the apparent discrepancy: it is the result of
different choices about which linear combination of Higgsino currents is used to
source the diffusion equations. References \cite{HN,AOS,CQRVW} considered only the
combination $H_1-H_2$, based on the observation \cite{HN} that the orthogonal source,
$H_1 + H_2$, is driven to zero in the limit that interaction rate $\Gamma$
proportional to the top quark Yukawa couplings becomes infinite.  Although this
approximation is convenient because it simplifies the network of coupled diffusion
equations, in the present application it can lead to a serious underestimate of the
BAU.  Parametrically, the $H_1+H_2$ source is suppressed by $(\Gamma D_{\tilde
h})^{-1/2}$, where $D_{\tilde h}\sim 20/T$ is the Higgsino diffusion constant
\cite{CJK2}. However for realistic values, $(\Gamma D_h)^{-1/2}\sim 1$, so there is
no actual suppression, eliminating the need for large CP violating phases to get the
observed BAU, which we had found with the $H_1-H_2$ source \cite{CJK}.  Here we will
demonstrate how this comes about, while at the same time updating our earlier
computation so as to give a quantitatively accurate determination of the BAU as a
function of relevant parameters in the MSSM.

\vspace{0.1in}

\noindent {\bf 2.}
To explain the how the classical force mechanism works, we first review
the simple case of a top quark with a spatially varying complex mass 
\cite{JPT}, $m(x) = yH(x)  e^{i\theta(x)}$.  By solving the Dirac equation 
$(i\dsl - |m|\cos\theta - i |m| \sin\theta\gamma_5)\psi$ to first 
order in derivatives (the WKB approximation), one finds that a particle 
of energy $E$ experiences a spin ($s$) dependent force
\beq
F = {dp\over dt}= -{(m^2)'\over 2 E} +
{s\over 2 E^2}\left({m^2\theta'	}\right)'.
\label{eq:force}
\eeq
The spin dependent part of the force has opposite sign for antiparticles,
which causes the distributions of like-helicity fermions and antifermions
to be separated in the vicinity of the wall.  For relativistic particles, 
we can approximate helicity by chirality, and speak of spatially varying 
chemical potentials, $\mu_{\sss L,R}$ for the left- and right-handed 
components; these are related to the asymmetry between the particle and 
antiparticle densities by $n(x) - \bar n(x) = \mu(x)T^2/6$.  Diffusion 
equations can be derived by inserting the force (\ref{eq:force}) into the 
Boltzmann equation, doing an expansion in moments of the distribution 
functions, and truncating the expansion. They have the form
\beq
-D\muL_i'' - v_w\muL_i' + \Gamma\muL_i = S_i(x),
\label{eq:diff1}
\eeq
where $D$ is the diffusion coefficient (of order the inverse mean free
path), $v_w$ is the bubble wall velocity, and $\Gamma$ is a damping rate
representing decays or inelastic collisions of the left-handed fermions.
The source term for each species is related to the classical force through a thermal
average \cite{CJK2}
\beq
S(x) = - {v_w D \over \avg{\vec v^{\,2}}} 
\avg{v_x F(x)}'. 
\label{eq:source}
\eeq
Although this looks similar to spontaneous baryogenesis \cite{SBG}, it is
not the same; the latter arises through the collision term in the Boltzmann equation,
while ours comes from the flow term \cite{CJK,CJK2}, and we find that it gives
a somewhat larger effect.

Once the left-handed quark density is found from eq.\ (\ref{eq:diff1}),
the density of baryons generated by sphalerons can be computed by
integrating $\muL(x) \equiv \sum_{q_i} \muL_i$ in front of the wall:
\beq
n_B = {9\Gsph T^2\over 2v_w}\int_0^\infty \muL(x)\, 
e^{-c_b\Gsph x/v_w}\,dx,
\label{eq:nB}
\eeq
where $\Gsph$ is the diffusion rate of the Chern-Simons number as measured 
by lattice simulations \cite{guy}: $\Gsph = (20 \pm 2)\alpha_w^5 T$. The 
exponential accounts for sphaleron-induced relaxation of the baryon 
asymmetry back to zero due to restoration of thermal equilibrium, in 
the limit of a very slowly moving wall.  The coefficient $c_b$ depends 
on the squark spectrum: it equals to $45/4$ if all squarks are heavy 
and $72/7$ if only the right handed stop is light.

\vspace{0.1in}

\noindent {\bf 3.} When the above picture is adapted to the MSSM some
complications occur, because the quarks do not get complex masses and
hence are not directly sourced.  The chargino mass matrix does contain
complex phases however, and the chiral asymmetry which develops in the
chargino sector induces one for the quarks because of the strong
coupling between Higgsinos and the top quark. This system involves not
just one, but many coupled diffusion equations. We will show that they
can, nevertheless, be reduced to a single equation by reasonable
approximations.

We start the treatment of the MSSM by deriving the source term analogous 
to $S(x)$ in eqs.\ (\ref{eq:diff1}--\ref{eq:source}). The mass term in the
Lagrangian for the charginos is 
\beq
  \overline\psi_R M_\chi \psi_L = (\overline{\widetilde w^{^+}},\
  \overline{\widetilde h^{^+}_{2}} )_{R}
  \left(\begin{array}{cc} 
             m_2 & g H_1 \\
           g H_2 & \mu 	
        \end{array}\right)
  \left(\begin{array}{c}
         \widetilde w^{^+} \\ 
         \widetilde h^{^+}_{1} 
        \end{array}
  \right)_{\!\!L}
\label{eq:chmass}
\eeq
plus the Hermitian conjugate, $\overline\psi_L M_\chi^\dagger \psi_R$. 
Because there are two Higgs doublets, $H_1$ and $H_2$, there are two
corresponding Higgsino fields $\widetilde h^{^+}_{1,2}$.  $\widetilde
w^{^+}$ is the wino, superpartner of the $W$ boson, and $g$ is the weak
gauge coupling.  The complex phases of the wino mass $m_2$ and the 
$\mu$ parameter are the origin of a CP violating force.  By again 
solving the Dirac equation in the WKB approximation, one can find the 
spin-dependent forces which act on the two mass eigenstates, with masses 
$m^2_\pm$.  Since Higgsinos couple strongly to top quarks, we are 
interested in the one which smoothly evolves into the Higgsino state in 
front of the wall, where $H_{1,2}\to 0$.  The spin-dependent part of the 
force has the same form as in eq.\ (\ref{eq:force}), with the spatially 
varying phase given by 
\beq
  m^2_\pm\theta_h' =  
        \mp {g^2{\rm Im}(m_2\mu)\over (m^2_+-m^2_-)}
        \left( H_1 H_2' + H_1' H_2 \right),
\label{eq:mssm_angle}
\eeq
where the Higgsino-like mass eigenvalue, $m^2_{\tilde h}(x)$, is $m^2_-$ 
(the lighter one)  if $|\mu|^2 < |m_2|^2$, and $m^2_+$ otherwise.  We 
remark that the effective WKB expansion parameter $\theta_h'/E$ remains 
small even when mass gap $m^2_+ - m^2_-$ attains its minimum value; this 
corresponds to $|m_2| \simeq |\mu|$, where one has parametrically 
$\theta_h'/E \simeq gH_i/wE|\mu| \ll 1$, because for typical wall 
widths $wE \gsim 20$ \cite{CMnote}.

Having specified the source term, we must now deal with the
diffusion equations in which it appears.  In general, these are a set
of coupled equations for the two Higgsinos ($\tilde h_{1,2}$), the
winos, 6 flavors of right-handed quarks, 3 generations of left-handed
quark doublets, and all the superpartners which are light enough to be
present at the temperature $T$.  Following Huet and Nelson \cite{HN},
we will make the simplifying assumption that the supergauge
interactions ({\it e.g.,} the coupling of winos to quarks and squarks)
are in thermal equilibrium, so that particle and corresponding
sparticle chemical potentials are equal to each other, species by
species; in particular gaugino chemical potentials are driven to zero.
We will further assume that the right-handed top squark is the only light
squark. The latter must be light to get a strong enough electroweak phase
transition, and it turns out that making all the others heavy is favorable
to electroweak baryogenesis, as well as satisfying constraints on CP
violation in the MSSM. 

With these simplifications, the diffusion equation network can be reduced
to four equations, for the potentials of the Higgsinos $\tilde h_1$, 
$\tilde h_2$ and the left-handed third generation quark doublet $q_3$ and 
the right-handed quark $t_R$.  Ignoring interactions involving left-handed 
squarks, which we assume to be heavy, the important interactions coupling 
these species come from the potential 
\beq
        V = y \bar q_3 H_2 t_R
        + \mu \tilde h_1 \tilde h_2 +{\rm h.c.}
\label{eq:pot}
\eeq
The rate for the Yukawa  interaction is $\Gamma_y$. 
In addition there are helicity 
flipping interactions with rate $\Gamma_{hf}$ coupling $\tilde h_1$ and 
$\tilde h_2$, due to the $\mu$ term, and in the broken phase inside the 
bubble ($x<0$), the top quark Yukawa coupling becomes the top mass, which 
causes helicity flips between the left- and right-handed top quarks, 
$q_3$ and $t_R$, with a rate of $\Gamma_m$.  Defining the diffusion 
operator ${\cal D}_i = -6(D_i \partial_x^2 + v_w\partial_x)$, and defining
also the convenient linear combinations $\mu_\pm = \frac12(\mu_{\tilde h_1} 
\pm \mu_{\tilde h_2})$ and 
$\mu_y = \mu_+ - \mu_- - \mu_{t_R} +\mu_{q_3}$, 
we have
\beqa
{\cal D}_h \mu_+  + 2 \Gamma_{hf}\mu_+ +\sfrac12 \Gamma_y \mu_y &=& S_H
\label{c1} \\
{\cal D}_h \mu_-  -\sfrac12 \Gamma_y \mu_y &=& 0
\\
\label{c2}
\sfrac13 {\cal D}_q \mu_{q_3}
                   + \sfrac12\kappa_{\tilde h}  \Gamma_y \mu_y
                   - \tilde \Gamma_m + 2 \tilde \Gamma_{ss} &=& 0
\\
\label{c3}
\sfrac12 {\cal D}_q \mu_{t_R}
                   -\sfrac12 \kappa_{\tilde h}\Gamma_y \mu_y
                  + \tilde \Gamma_m - \tilde \Gamma_{ss} &=& 0.
\label{c4}
\eeqa
In deriving these equations we have assumed that all squarks but right
handed decouple from thermal equilibrium. The factor $\kappa_{\tilde h} 
\simeq \sfrac12 x_{\tilde h}^2 K_2(x_{\tilde h})$, where $x_{\tilde h}=
m_{\tilde h}/T$ and $K_2$ is the modified Bessel function, accounts for 
partial decoupling of higgsinos.  The top-quark helicity-flip term is defined 
to be $ \tilde \Gamma_m  \equiv \Gamma_m \theta(-x) (\mu_{t_R}-\mu_{q_3})/T$
and that for the strong sphaleron rate is 
$\tilde\Gamma_{ss} \equiv \Gamma_{ss}( 20\mu_{q_3}+ 26\mu_{t_R})/T$
with $\Gamma_{ss} = 1500 \kappa_{\rm sph} \alpha_W^5 T$ \cite{Ssguy}.

It is noteworthy that only the linear combination $\mu_+$ gets directly
sourced in our WKB treatment, whereas the source for the combination 
$\mu_-$ vanishes (in this respect ref.\ \cite{CJK} was in error).  This 
is in contrast to papers which  treat the particle reflections from the 
wall quantum mechanically; these works find that both $\mu_+$ and $\mu_-$ 
are sourced.  Here we point out that, within the quantum reflection 
formalisms, it is still true that $\mu_+$ has a larger source than 
$\mu_-$, because $H_1'H_2 + H_2'H_1$ is larger than $H_1'H_2 - H_2'H_1$ 
\cite{Rius_note}.  However nobody has heretofore considered the former 
because of the unfortunate approximation of imposing equilibrium of the 
Yukawa interactions $(\Gamma_y\to\infty)$, which forces $\mu_+$ to zero.  

The network of diffusion equations can be integrated numerically to to
obtain $\mu_{t_R}$ and $\mu_{q_3}$.  From these one can compute the
left-handed quark potential $\mu_L = 18\mu_{t_R} + 15\mu_{q_3}$, which must
then be numerically integrated in eq.\ (\ref{eq:nB}) to obtain the  baryon
asymmetry.  

\vspace{0.1in}

\noindent {\bf 4.}
We have carried out the above procedure to study how the BAU depends on
the velocity of the bubble wall $v_w$, the wall width $w$ (appearing in
the Higgs field profile as $H_0(1-\tanh(x/w))/2$), $\tan\beta = 
\avg{H_2}/\avg{H_1}$ (at zero temperature) and the chargino mass parameters 
$\mu$ and $m_2$.  We take as our preferred values $\tan\beta = 3$, $w = 6/T$ 
\cite{CMnote}, $v_w = 0.1$ and $\mu = m_2 = 100$ GeV. We also took 
$T=90$ GeV and $H_0 = 110$ GeV. We computed the values of $D_{\tilde h}$, 
$\Gamma_y$, $\Gamma_{hf}$ and $\Gamma_m$ from the actual Feynman diagrams 
and thermally averaged cross sections, and also using the results of ref.\ 
\cite{AMY}. With our preferred parameters we find $D_{\tilde h} = 80/T$, 
$\Gamma_m = 0.0012T$, 
$\Gamma_{hf} = 0.001 x_{\tilde h}^2/[(1+x_{\tilde h})(2+x_{\tilde h})]$ 
and 
$\Gamma_y = [0.012 + 0.022 \,x_h \,(1-1.87/x_h^{2}+ 0.05/x_h^{4})^{3/2}]T$.  
In the following figures where certain of 
these quantities are varied, those which are not specified have the above
values.  It is customary to express the BAU as 
a scaled ratio of baryons to photons, $\eta_{10} = 10^{10} n_B/n_\gamma 
= 7\times 10^{10}n_B/s$, where $s =(2\pi^2/45)g_*T^3$, and we take the 
number of degrees of freedom in the MSSM to be $g_* = 110$.  Current 
limits from big bang nucleosynthesis give $3\lsim\eta_{10}\lsim 4$.

In figure 1, the BAU is plotted as a function of wall velocity, varying
the width of the bubble wall to obtain the different curves.  We have 
taken the CP-violating phase appearing in Im($m_2\tilde\mu$) (eq.\ 
(\ref{eq:mssm_angle})) to be maximal; to satisfy $\eta_{10}\cong 3$, one 
should rescale this phase accordingly. The efficiency of baryogenesis tends 
to peak for $v_w= 0.01-0.1$.  Interestingly, recent work on bubble expansion 
has suggested just such a range of small values in the MSSM\cite{vel}.  

\vspace{0.02in}
\begin{figure}[h]
\centerline{\epsfxsize=3.0in\epsfbox{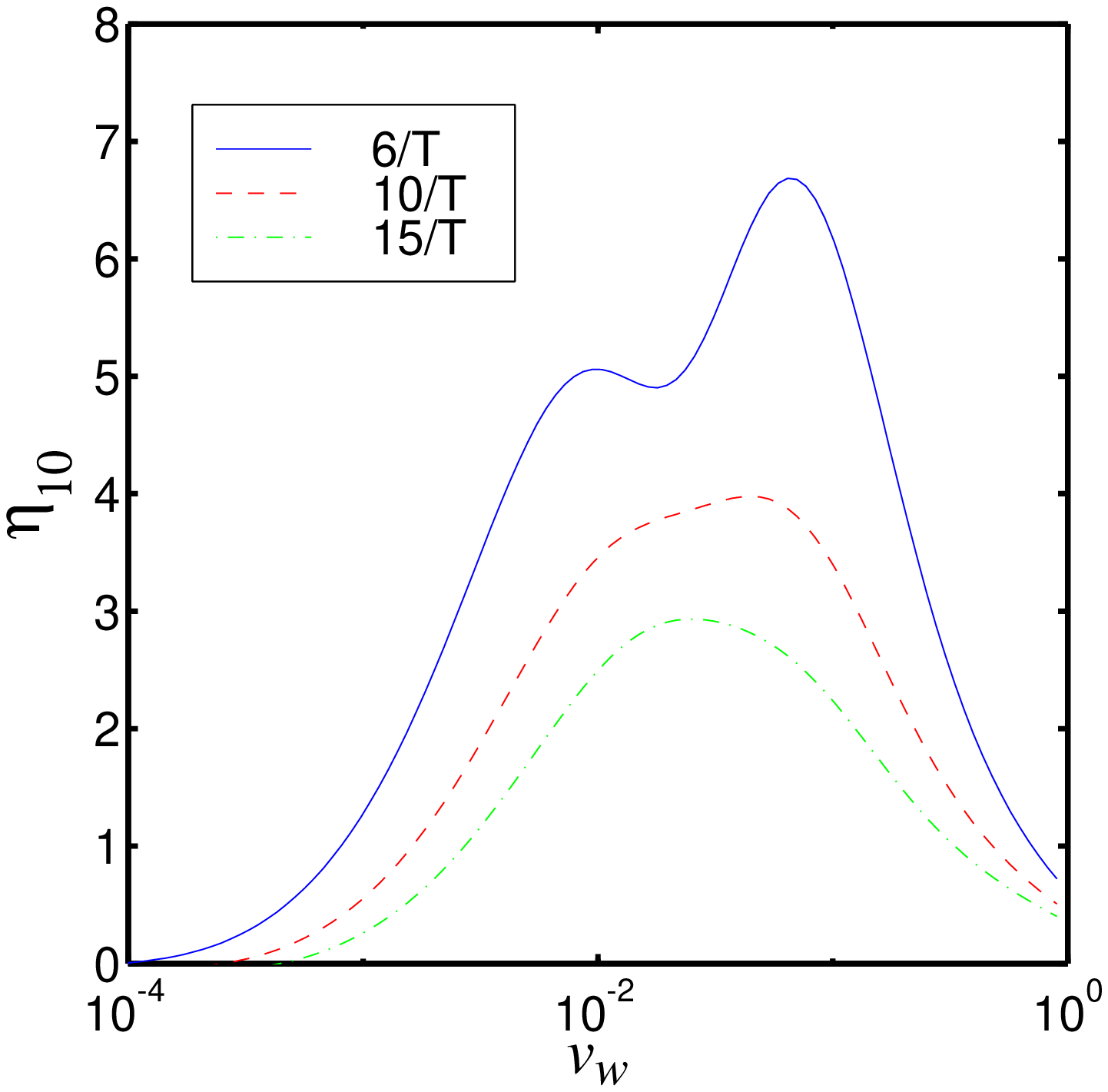}}
\vspace{-0.03in}
\noindent
{\small Figure 1: Baryon asymmetry $\eta_{10}$ as a function of wall 
velocity $v_w$ for wall width $w = 6/T$, $10/T$ and $15/T$.}
\end{figure}
\vspace {0.05in}

We can also consider the dependence of the BAU on the chargino mass
parameters, $\mu$ and $m_2$.  Figure 2 exhibits contours of
constant $\eta_{10}$ assuming the maximal  magnitude of the CP-violating
phase, $\arg(m_2\mu) = \pi/2$, and wall velocity $v_w=0.1$. Since we need
$\eta_{10}\cong 3$, and the maximum value in the figure occurs around
$\eta_{10}\cong 20$ for $\mu\sim m_2\sim 50$ GeV, we find that CP phase
can be no smaller than 0.15, 
and then only if the charginos 
are rather light. Such large values of the CP phase could come into
conflict with the most stringent present experimental constraints, which
come from searches for the electric dipole moment of the $^{199}$Hg atom
\cite{FOPR}.  In order to evade these constraints, one must assume that
most of the squarks or gauginos contributing to EDM loop diagrams are
heavy, in the TeV range.

\begin{figure}[h]
\centerline{\epsfxsize=3.5in\epsfbox{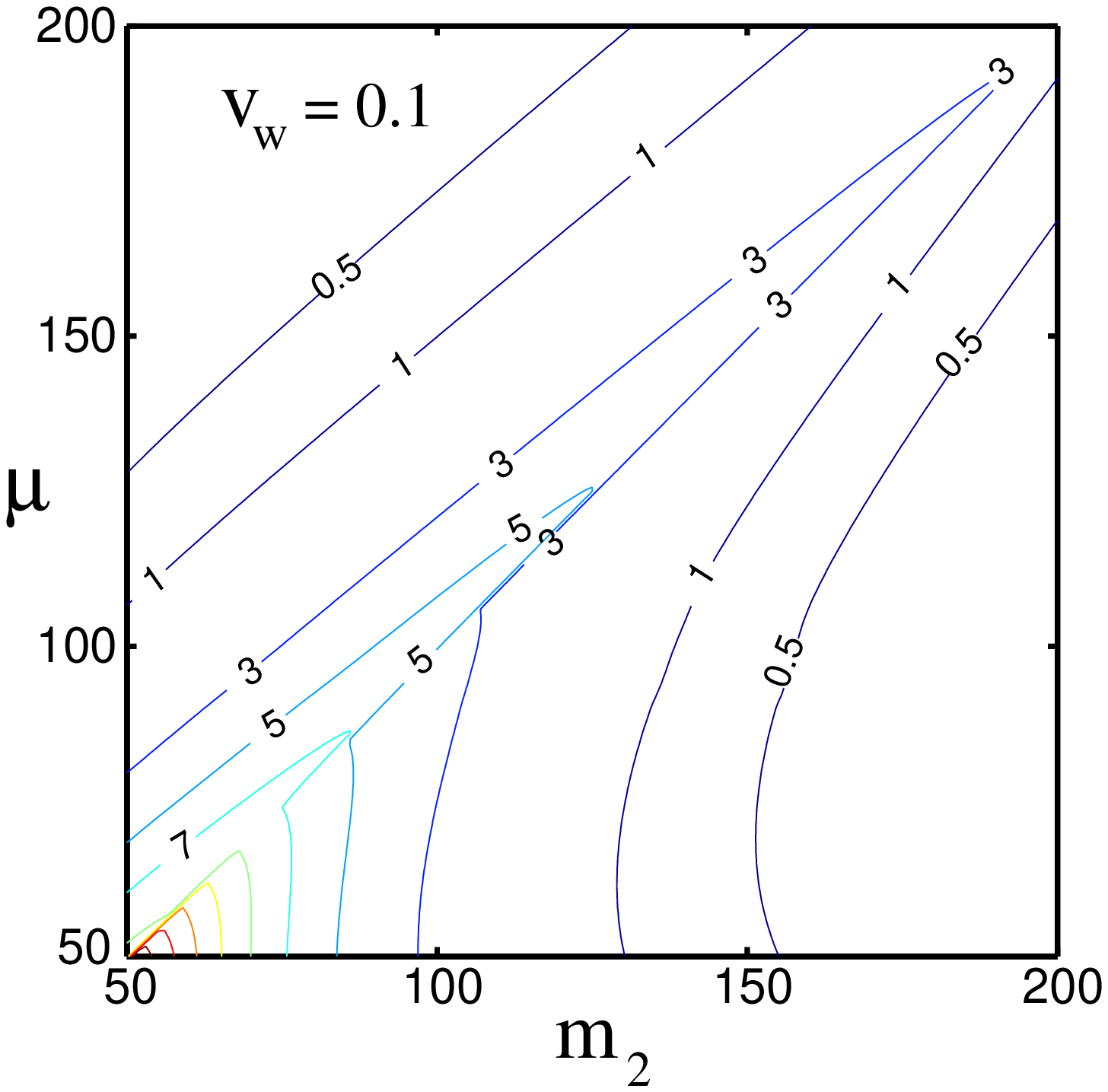}}
\vspace{-0.03in}
\noindent
{\small Figure 2: Contours of constant $\eta_{10}$, assuming maximal
CP violation ($\arg(m_2\mu)=\pi/2$) and $v_w=0.1$, 
in the plane of the chargino mass
parameters $\mu$ and $m_2$ (in GeV units).}
\end{figure}
\vspace{-0.0in}

\vspace{0.1in}

\noindent {\bf 5.} In summary, we have presented a quantitatively
accurate analysis of the baryon asymmetry of the universe in the MSSM,
using the classical force mechanism, which allows a consistent
computation of the sources appearing in the diffusion equations.  A
CP-violating force acts on Higgsinos while they cross the bubble walls
causing a particle-antiparticle separation in charginos, which then
gets partially transformed to a chiral quark asymmetry that biases
sphalerons to produce baryons. Unlike previous authors, we have not
assumed that the two components of the Higgsino, $\tilde h_1$ and
$\tilde h_2$, reach complete chemical equilibrium through Yukawa and
helicity-flipping interactions, and we showed that a large enhancement
of the BAU can result, relative to what one would get this
oversimplification.  Nevertheless, after carefully computing the relevant
damping rates and diffusion coefficients, and accurately solving the
transport equations, we find that it is difficult to generate the observed
baryon asymmetry unless CP violation in the $\mu$ term is close to
maximal.    Further details of our formalism and
calculations can be found in ref.\ \cite{CJK2}.

\vspace{0.1in}

We thank Guy Moore for very helpful correspondence, and in particular Michael 
Joyce and Tomislav Prokopec for collaboration on related issues. Special thanks
also to Stephan Huber and Michael Schmidt for pointing out an error in our 
original numerical results. KK thanks  CERN for hospitality during the 
completion of this work.


\begin{references}
\bibitem{CMnote} 
  The simulations of the electroweak phase transition done in ref.\ 
  \cite{CM} indicate that the wall width in the MSSM is $w = (10\pm 4)/T$.
\bibitem{Rius_note} 
  Ref.\ \cite{Rius}, working in the thin wall limit, notes that 
  $\mu_-$ does get an unsuppressed source at next order ($O(H^4/T^4)$) in an
  expansion in powers of the background Higgs fields.
\end{references}
\end{document}